\documentclass[final,3p,times,twocolumn]{elsarticle}

\usepackage{graphicx}
\usepackage[colorinlistoftodos]{todonotes}
\usepackage{amssymb}
\usepackage{braket}

\usepackage{xspace}

\biboptions{comma,sort&compress}

\usepackage{amsmath}
\usepackage{siunitx}
\usepackage{cleveref}
\usepackage{CJKutf8}
\usepackage{rotating}
\usepackage{booktabs}
\usepackage{subfig}
\usepackage[version=4]{mhchem}
\journal{Journal of Molecular Spectroscopy}

\usepackage{xcolor,textcomp}

\begin{document}

\begin{frontmatter}

\title{Millimeter-millimeter-wave double-modulation double-resonance spectroscopy}

\author[ph1]{Oliver Zingsheim\corref{cor1}}
\ead{zingsheim@ph1.uni-koeln.de}
\author[ph1]{Luis Bonah}
\author[ph1]{Frank Lewen}
\author[ph1]{Sven Thorwirth}
\author[ph1]{Holger S. P. M\"uller}
\author[ph1]{Stephan Schlemmer}
\cortext[cor1]{Corresponding author}

\address[ph1]{I. Physikalisches Institut, Universit\"at zu K\"oln, Z\"ulpicher Stra{\ss}e 77, 50937 K\"oln, Germany}

\begin{abstract}

A new millimeter- to millimeter-wave double-modulation double-resonance (MMW-MMW DM-DR) scheme has been applied to record spectra of two astronomically relevant complex organic molecules (COMs), propanal (\ce{C2H5CHO}) and ethyl cyanide (\ce{C2H5CN}), to demonstrate advantages of the DM-DR experimental technique. 
The DR technique helps to identify target transitions in a forest of lines and the implementation of a DM procedure (modulation of the pump and probe source) allows for confusion- and baseline-free spectra containing only the line(s) of interest.
In particular the unambiguous assignment of weak and blended transitions in very dense MMW spectra is highlighted.
Details of the observed Autler-Townes line splitting and possible future applications, such as automated analyses and adaptions of DM-DR methods to other experimental setups, are discussed.

\end{abstract}


\end{frontmatter}

\section{Introduction}                        
\label{Sec:Introduction}                      

The plethora of lines observed in the microwave (MW) and (sub-)millimeter-wave (sub-MMW) spectra of complex organic molecules (COMs) often originate from the simultaneous presence of different conformers, their various isotopologues and of energetically low-lying vibrationally excited states.
Analyses of these spectra may be complicated further by vibration-rotation or other interactions, see, e.g., the recent analyses of acrylonitrile (\ce{CH2CHCN}) \cite{KISIEL201583}, propanal (\ce{CH3CH2CHO}) \cite{ZINGSHEIM2017_Propanal}, ethyl cyanide (\ce{CH3CH2CN}) \cite{KISIEL2020111324,ENDRES2021_EtCN}, and references therein.
Unambiguous spectroscopic assignments in the (sub-)MMW region are essential for deriving accurate descriptions of complex spectroscopic molecular fingerprints, eventually allowing for the identification of molecules in space.
Approaches simplifying the analysis are therefore highly desirable \cite{Fortman_2010}.
Double-resonance (DR) and double-modulation (DM) are well-suited experimental techniques in this regard, as they help to identify specific transitions and clear up molecular spectra, respectively.

DR experiments may play a crucial role in finding spectroscopic linkages and with that to assign unambiguously quantum numbers to observed transitions.
This is especially important for perturbed transitions, see, e.g., Refs.~\cite{1995JMoSp.172...57C,2001JMoSp.205..185C,Pate2010_Coherence-converted}.
DR is also used to extend the frequency region of a given setup, e.g. to extend Fourier transform microwave (FTMW) spectrometers to MMW \cite{Jager_Gerry_1995,Suma_Endo_2004}, or to populate states of  higher energy \cite{1981JMoSp..90..222J}. 
DR techniques are well-established, in particular in MW-MW setups
\cite{Crabtree2016_Taxonomy,Martin-Drumel2016_AMDOR}, in MW-MMW \cite{1995JMoSp.172...57C,2001JMoSp.205..185C,Jager_Gerry_1995,Suma_Endo_2004,Roenitz_2018}, and in various other frequency regions (a collection of experimental results spanning various frequency regions is given in Ref.~\cite{Roenitz_2018}; references cited here are just a selection of numerous DR related publications).
Finding linkages between two (the "pump" and "probe") transitions which share an energy level is enabled by monitoring the intensity alteration of a probe signal. This is often done by inducing population changes \cite{Jager_vanderWaals_1998}, or by destroying the phase coherence in chirped pulse (CP) experiments \cite{Suma_Endo_2004,Crabtree2016_Taxonomy}.
Furthermore, 
the Autler-Townes effect \cite{AutlerTownes1955,Cohen-Tannoudji2008_DressedAtomApproach} may explain DR phenomena \cite{Schmitz-Patterson-Schnell_2015,Roenitz_2018}, which is particularly true for results presented in this paper.

The DM technique results in confusion- and baseline-free spectra containing only the line(s) of interest.
These experimental advantages allow for the measurements of weaker signals and have already been demonstrated in the literature, e.g. Refs.~\cite{AMANO1974,AMANO2000_DM_HOC,Gudeman_Saykally_1983,Havenith1994_DM_concentration_Ar-CO}.
For example, frequency modulation (FM) of the probe source was used in combination with Zeeman modulation to detect weak lines of vibrationally excited \ce{O2} \cite{AMANO1974}.
Moreover, 
in combination with a glow discharge, the detection of weak signals of the \ce{HOC+} ion was facilitated \cite{AMANO2000_DM_HOC}.
Velocity-modulation spectroscopy allowed to measure ro-vibrational transitions of \ce{HCO+}, by modulating the electrode potentials in a glow discharge and therefore suppressing the signals of the much more abundant neutral (precursor) species \cite{Gudeman_Saykally_1983}. 
Furthermore, a concentration-frequency double-modulation scheme, implemented using a discharge within the expansion region of a jet, was used to measure weak signals of complexes in the infrared region, in particular to get rid of signal fringes \cite{Havenith1994_DM_concentration_Ar-CO}. 

In this paper, DR and DM experimental techniques are combined in the MMW region to unambiguously assign weak rotational transitions in a forest of lines.
Based on the fundamental importance of its effect to experimental results presented here, the already well-understood Autler-Townes effect and its influence to different 3-level system arrangements is revisited next (Sec.~\ref{Sec:Autler-Townes}).
Then, the novel MMW-MMW DM-DR experimental scheme, DM as both the probe and pump sources are modulated, is presented (Sec.~\ref{Sec:DR Experiment}), including a detailed description of observed phenomena based on the Autler-Townes effect.
In the following, first DM-DR measurements of COMs are presented, highlighting in particular unambiguous assignments of transitions 
with a special reference to very weak and blended transitions (Sec.~\ref{Sec:ExperimentalResults}).
Subsequently, the measurements are discussed and conclusions are drawn (Sec.~\ref{Sec:Discussion}).
Finally, an outlook on future applications, in particular extending established experimental setups with the DM-DR technique and automated analyses, are highlighted (Sec.~\ref{Sec:Outlook}).

\bigskip

\section{Autler-Townes effect}                  
\label{Sec:Autler-Townes}

\begin{figure}[t]
\centering
\includegraphics[width=\linewidth]{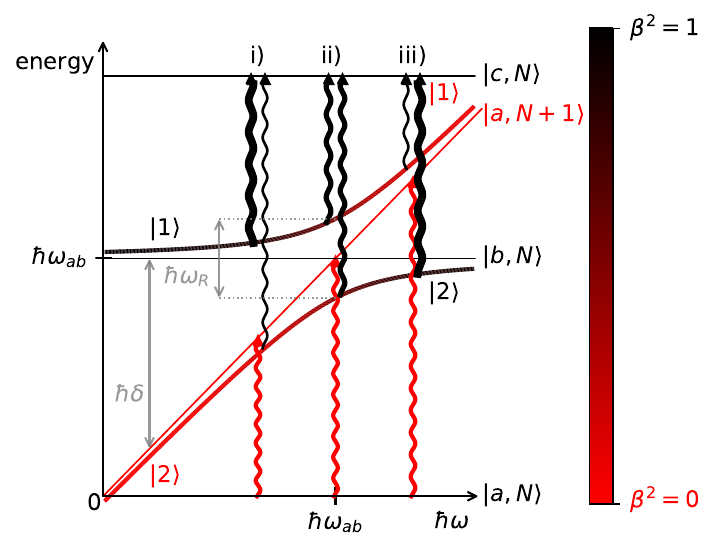}
\caption{
Simplified Autler-Townes effect of a progressive 3-level system when a 2-level system is pumped.
The probe transition is split into two components (black curly lines) when the pump radiation $\hbar \omega$ is intense enough (red curly lines) to split the dressed states $\ket{1}$ and $\ket{2}$ prominently.
This splitting is highlighted for three different cases: i) red detuned pumping with $\omega<\omega_{ab}$ ($\delta<0$), ii) on-resonance pumping with $\omega=\omega_{ab}$ ($\delta=0$), iii) blue detuned pumping with $\omega>\omega_{ab}$ ($\delta>0$).
The intensity of the probe transitions is depicted by the thickness of the curly black arrows and is proportional to the contributing mixing of $\ket{b,N}$, $\beta^2$, of the respective dressed state $\ket{1}$ or  $\ket{2}$.
The figure is adapted from Ref.~\cite{Cohen-Tannoudji2008_DressedAtomApproach}.
}
\label{FigA:Autler_Townes}
\end{figure}

In a semi-classical picture, Rabi oscillations of a molecular two-level system, $\ket{a}\leftrightarrow \ket{b}$, with energy difference $E_b-E_a=\hbar\omega_{ab}$, occur if a pump radiation close to its resonance frequency $\omega_{ab}$ is applied. The Rabi frequency
 \begin{equation}
 \omega_R=\frac{\mu_{ab}E}{\hbar} 
 \label{Eq:Rabi_dip+E}
 \end{equation}
depends on the transition dipole moment $\mu_{ab}$ and the applied electric Field intensity $E$ and causes population changes. 
This may be observed in the resulting intensity alteration of a second, probe transition which shares one energy level with the pumped system, e.g. $\ket{b}\leftrightarrow \ket{c}$.
In 1955, Autler and Townes observed an additional splitting of probed transitions in OCS molecules if an intense pump radiation was applied \cite{AutlerTownes1955}, which can not be explained by means of a semi-classical picture and which is nowadays referred to as the Autler-Townes effect~\cite{Cohen-Tannoudji2008_DressedAtomApproach}. The description of this splitting is summarized in the following based on the literature as it is of fundamental importance to interpret the results of this paper. 

The Autler-Townes splitting can only be explained in a full quantum mechanical picture, see Fig.~\ref{FigA:Autler_Townes}.
A 3-level arrangement is considered initially in which the two allowed transitions, pump and probe, connect the three molecular energy levels with energies $E_c>E_b>E_a$ in a ladder fashion (commonly referred to as a \textit{progressive} arrangement).
In a first step, only the pumped 2-level system is considered for further clarification ($\ket{a,N}\leftrightarrow \ket{b,N}$).
This time, in contrast to the semi-classical picture, the two pumped energy levels are each dressed with $N$ photons when a powerful pump radiation, $\hbar\omega$, is applied (cf. red curly arrows).
The two unperturbed dressed states are parallel to each other as they are dressed with the same number of photons, such that increasing the pump frequency $\omega$ ($x$-axis) effects both levels in the same way. 
$\ket{a,N}$ may be arbitrarily set to 0 and the energy difference to $\ket{b,N}$ is $\hbar\omega_{ab}$.
The dressed state $\ket{a,N+1}$ linearly depends on the pump photon frequency $\omega$ and has exactly the same energy as $\ket{b,N}$ for zero detuning, $\delta=\omega-\omega_{ab}=0$, or also called on-resonance pumping ($\delta=0\Rightarrow E_{a,N+1}=E_a+\hbar\omega=E_a+\hbar\omega_{ab}=E_{b,N}$).
 
The two levels $\ket{b,N}$ and $\ket{a,N+1}$ will interact considerably if they are close in energy ($\delta\approx0$). 
The energy splitting of the resulting perturbed dressed states, $\ket{1}$ and $\ket{2}$, depends on the detuning $\delta$ and on the Rabi frequency $\omega_R$ and is given by
\begin{equation}
    \Delta E =\hbar  \sqrt{\omega^2_R+\delta^2}.
\label{Eq1:Autler-Townes_splitting}
\end{equation}
If the system is pumped on-resonance ($\delta=0$), resulting in $\Delta E= \hbar \omega_R=h\nu_R$, being case ii) in Fig.~\ref{FigA:Autler_Townes}, the frequency shift of the two components (or the Rabi frequency in Hz)
\begin{equation}
\nu_R=\frac{\mu_{ab}}{h}\sqrt{\frac{8P}{\pi d^2\epsilon_0c}}
\label{Eq2:Autler-Townes_Rabi-frequency}
\end{equation}
can be calculated with the output power of the pump radiation $P$, its beam size diameter $d$, and the transition dipole moment $\mu_{ab}$ and is obtained by using various basic relations \cite{Bernath}, additionally presented in Eq.~\eqref{Eq:AT_power_formula} in the Appendix. 
For example, a transition dipole moment of $\mu_{ab}=1$\,D and a pump radiation power of $P=10$\,mW with a beam size diameter of $d=10$\,cm results in a splitting of the two energy levels of about $\nu_R=156$\,kHz.

Instead of a single transition, $\ket{b}\leftrightarrow \ket{c}$, \textbf{the probe transition splits into two components, one red and one blue shifted one, as two transitions, $\ket{1}\leftrightarrow \ket{c,N}$ and $\ket{2}\leftrightarrow \ket{c,N}$ are allowed}, see black curly arrows in Fig.~\ref{FigA:Autler_Townes}.
The energies of the perturbed levels, whose splitting is described by Eq.~\eqref{Eq1:Autler-Townes_splitting}, form hyperbolas to the straight lines of the unperturbed states $\ket{a,N+1}$ and $\ket{b,N}$ in Fig.~\ref{FigA:Autler_Townes}.
This so-called avoided crossing behavior can be studied experimentally by systematic detuning of the pump frequency while the two probe components are scanned.
This may be visualized conveniently in 2D pump-probe DR spectra.
Furthermore, considering the third possible transition to be forbidden, $\ket{a,N}\nleftrightarrow \ket{c,N}$, the two probe components, $\ket{i}\leftrightarrow\ket{c,N}$, are more intense if the perturbed levels $\ket{i}$ are energetically closer to $\ket{b,N}$.
Mathematically, the intensity of each component depends on $\beta$ considering a mixing described by $\ket{i}=\alpha\ket{a,N+1}+\beta\ket{b,N}$ with $\alpha^2+\beta^2=1$ and $i=1,2$, where the mixing coefficients, or $\alpha$ and $\beta$, are described by $sin(\Theta)$ or $cos(\Theta)$ terms with the relation $tan(2\Theta)=-\omega_R/\delta$ with $0\leq2\Theta<\pi$.
The intensity relation of the two split probe components, $\ket{1}\leftrightarrow \ket{c,N}$ and $\ket{2}\leftrightarrow \ket{c,N}$, is given by
\begin{equation}
\left(\frac{I_{1}}{I_{2}}\right)^{\pm 1}=\tan^2\left[ \frac{1}{2}\arctan\left(-\frac{\omega_r}{\delta}\right)  \right]=\frac{I_{a,N+1}}{I_{b,N}}
\label{Eq3:Autler-Townes_Intensity}
\end{equation}
with the power of $+1$ for $\delta>0$ or $-1$ for $\delta<0$.
At $\delta=0$ a so-called label switching occurs, in case $\ket{1}$ or $\ket{2}$ are labeled either with $\ket{a,N+1}$ or $\ket{b,N}$ in regard of the higher mixing contribution $\alpha>\beta$ or $\alpha<\beta$, respectively.  
Equally strong intensities of the two components for zero detuning are observed ($\beta^2=0.5$; case ii in Fig.~\ref{FigA:Autler_Townes}), though, the red shifted component is more intense for red detuned pump frequencies, whereas, the blue shifted component is more intense for blue detuned pumping. However, this only holds for progressive energy level arrangements, such as depicted in Fig.~\ref{FigA:Autler_Townes}. 

\begin{figure}[t]
\centering
\includegraphics[width=\linewidth]{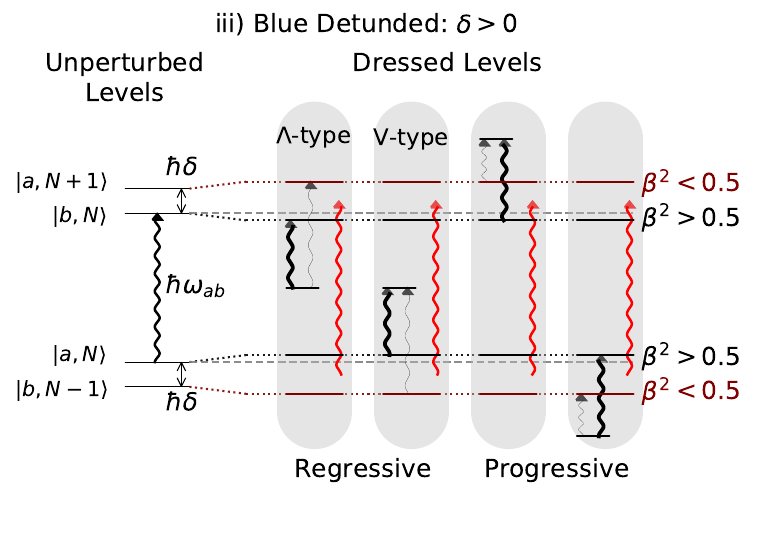}
\caption{
The four possible 3-level energy level arrangements of a blue detuned pumped 2-level system ($\delta>0$, case iii of Fig.~\ref{FigA:Autler_Townes}).
The red-shifted component of the probe transition is more intense for regressive energy level arrangements, whereas the blue-shifted component for progressive ones. Vice versa if the pump frequency is red detuned.
}
\label{FigA:Autler_arrangements}
\end{figure}

There are in fact four possible 3-level arrangements if a certain two-level system is pumped. Two progressive arrangements, energy levels are connected in a ladder fashion, and two regressive ones, where pump and probe transitions end ($\Lambda$-type) or start ($V$-type) in one unperturbed energy level, see Fig.~\ref{FigA:Autler_arrangements}. So far it was neglected that also the dressed state $\ket{b,N-1}$ has the same energy as $\ket{a,N}$ for on-resonance pumping.
Therefore, both energy levels of the pumped 2-level system, $\ket{a,N}$ and $\ket{b,N}$, will split if the pump radiation is powerful enough and the detuning is small. 
In case of a blue detuned pump frequency, the energy level $\ket{b,N-1}$ is lower in energy than $\ket{a,N}$, whereas $\ket{a,N+1}$ is higher than $\ket{b,N}$.
Considering different $\beta^2$'s of the perturbed levels, intensity ratios of the two probed Autler-Townes components depend on the level arrangement, cf. Eq.~\eqref{Eq3:Autler-Townes_Intensity}.
For a blue detuned pump radiation ($\delta>0$), the red-shifted component of the probe transition is more intense for regressive energy level arrangements, whereas the blue-shifted component for progressive ones, see Fig.~\ref{FigA:Autler_arrangements}.
On the other hand, the blue-shifted component of the probe transition is more intense for regressive energy level arrangements, whereas the red-shifted component for progressive ones if the pump radiation is red detuned ($\delta<0$).
The frequency shift of the more intense component is often referred to as the AC Stark shift, as in extreme cases only the more intense component is detectable which is then shifted compared to the unperturbed center frequency.
Differing intensity ratios are the basis to distinguish regressive and progressive energy level arrangements of 3-level systems in DR rotational spectroscopy \cite{Schmitz-Patterson-Schnell_2015,Roenitz_2018}.
For example, in CP-FTMW experiments, \citet{Schmitz-Patterson-Schnell_2015} showed that the Autler-Townes splitting causes a 180$^{\circ}$ phase change between progressive and regressive arrangements.
Furthermore, in MW-MMW DR measurements, regressive and progressive arrangements were distinguished by slightly detuned pump frequencies as has been described before \cite{Roenitz_2018}.

Expected Autler-Townes splittings are on the order of 100\,kHz which is in agreement with typical line widths of COMs in the MMW region.
Most important to the results of this paper is that observable splittings produced by commercially available amplifiers are large enough to induce clear intensity changes of the probed transition.

\bigskip
\section{Experimental Setup}                  
\label{Sec:DR Experiment}

\begin{figure}[t]
\centering
\includegraphics[width=1.0\linewidth]{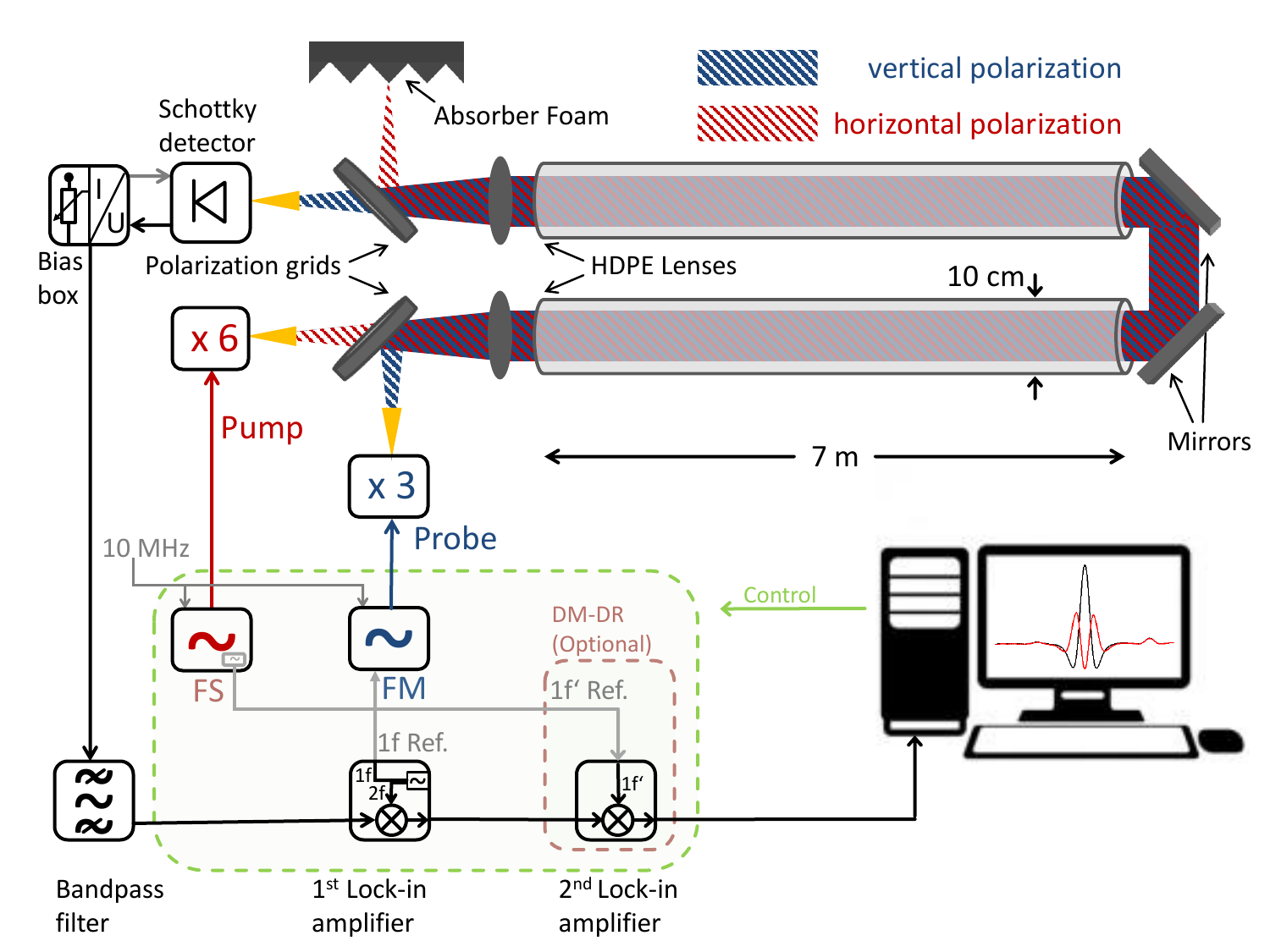} 
\caption{MMW-MMW DM-DR experimental setup. Probe (in blue) and pump (in red) beams are orthogonally polarized to each other.
The signal path is shown by black arrows, other electronic signal paths are shown in grey. The signal generator and the second lock-in amplifier are only used for DM experiments. For more details see text.}
\label{Fig:exp_setup_schematic}
\end{figure}

(Sub-)MMW absorption spectroscopy of molecular rotational transitions is usually conducted using FM to increase the signal-to-noise ratio (SNR) of measurements \cite{PhysRevLett.25.1397,MEDVEDEV2004314,ZAKHARENKO201541_FM_Lille,Esposti_2017_2f_Bologna,Motoki_2013_FM_Toho,Ordu2019_Acetone,Drumel2015_OSSO}.
For DR and DM-DR experiments, in addition to the probe radiation, a pump radiation is applied which might  (DM-DR) or might not (DR) be modulated. A schematic of the experimental setup capable of DM-DR measurements is shown in Fig.~\ref{Fig:exp_setup_schematic}.
First, the ``conventional" spectrometer is introduced, which allows to measure the rotational fingerprints in the MMW region (Sec.~\ref{subSec:Convenional Experiment}).
The setup complemented with a second radiation source for DR experiments to identify linkages between transitions is elucidated afterwards with a focus on the description of observed phenomena based on the Autler-Townes effect (Sec.~\ref{subSec:DR Experiment}).
Finally, a detailed description of the operation principle of DM-DR measurements is presented which allows securely to assign weak and blended transitions as spectra are greatly simplified (Sec.~\ref{subSec:DM-DR Experiment}).

\subsection{Conventional Setup - Measure Rotational Lines}                    
\label{subSec:Convenional Experiment}                  

Initially, the probe beam is considered only (in blue, Fig.~\ref{Fig:exp_setup_schematic}), whereas the pump beam (in red; DR off) is not applied yet and the optional lock-in amplifier is not used.
A conventional experimental setup consists in general of three major parts, i) frequency source, ii) absorption cell and iii) detector.
An Agilent E8257D synthesizer is used as (probe-)frequency source and referenced to a 10\,MHz rubidium atomic clock providing $\Delta\nu/\nu=10^{-11}$.
The probe signal is frequency tripled and electronics developed in-house are used for operation in full saturation mode to ensure a stable output power of about 1\,mW (0\,dBm). 
The frequency range covered is 70 to 129\,GHz.
The lock-in amplifier submits a reference frequency for the FM ($f\approx47$\,kHz). The amplitude of the FM is usually on the order of the full width at half maximum (FWHM) of the absorption profile.  
The frequency modulated radiation is sent through the cell via a gold-coated horn antenna. Polarizing grids, high-density polyethylene (HDPE) lenses with 15\,cm focal length, and planar mirrors are used to guide the probe beam (plane waves with an assumed beamsize diameter of 10\,cm) through the cell.
The cell and the windows are made of Pyrex (borosilicate glass) and Teflon (polytetrafluoroethylene; PTFE), respectively.
The effective absorption path length is 14\,m ($=2\times7$\,m).
The received signal of the Schottky detector is amplified in an in-house developed bias-box and then processed by a bandpass-filter.
The signal is demodulated finally with the 2$f$ reference signal in the lock-in amplifier. The 2$f$ demodulation mode creates absorption features that are close to the second derivative of a Voigt-profile, see black trace in Fig.~\ref{Fig:DR_singleLine}.

\subsection{DR Setup - Identifying Linkages}                    
\label{subSec:DR Experiment}                  

\begin{figure}[t!]
\centering
$\vcenter{\hbox{\includegraphics[width=0.65\linewidth]{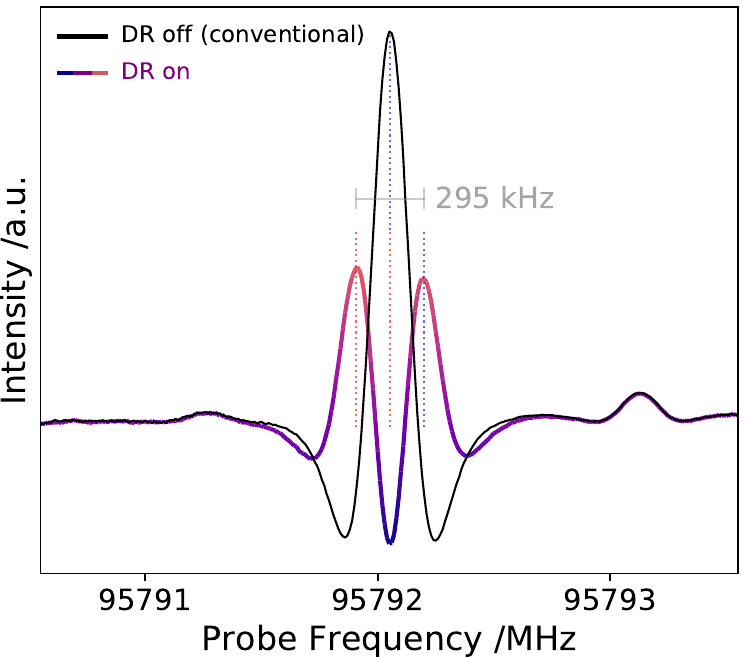}}}$
$\vcenter{\hbox{\includegraphics[width=0.32\linewidth]{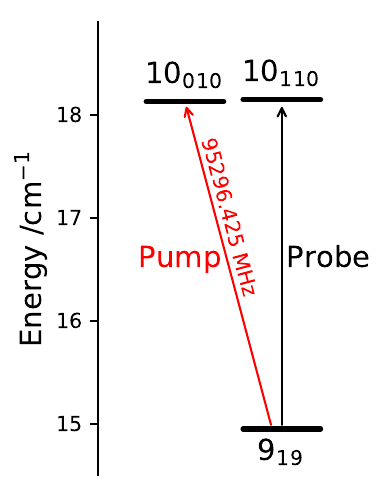}}}$
\caption{Conventional (in black) and DR measurement (in purplish) of propanal.
The splitting of the red- and blue-shifted Autler-Townes components 
of the $J_{K_a,Kc}=10_{1,10}\leftarrow 9_{1,9}$ $a$-type transition 
is 295\,kHz, when the $J_{K_a,Kc}=10_{0,10}\leftarrow 9_{1,9}$ $b$-type transition is pumped on-resonance 
(regressive energy ladder arrangement, cf. Fig.~\ref{FigA:Autler_arrangements}).
}
\label{Fig:DR_singleLine}
\end{figure}

The additional key component of the DR setup is a Rohde \& Schwarz SMF100A synthesizer which is used as pump source to study 3-level systems described in Sec.~\ref{Sec:Autler-Townes}.
The optional lock-in amplifier is still not yet used because the frequency of the pump source is unaltered, i.e. not yet modulated.
An active frequency multiplier (AFM6 70-110+14 from RPG Radiometer Physics) is used as pump source delivering output powers of up to 60\,mW (17.8\,dBm).
The final pump and probe frequencies are both located in the W-band. Plane waves are again assumed to pass the absorption cell with a beamsize diameter of $d\approx10$\,cm.
The polarization of the pump radiation is orthogonal to the probe radiation.
The Schottky detector is protected in this way from the high output power of the pump source.

If pump and probe frequencies are resonant to two molecular transitions which share an energy level, the probe intensity is altered due to the pump radiation, which is demonstrated in Fig.~\ref{Fig:DR_singleLine}.
The splitting of the two Autler-Townes components of the $J_{K_a,Kc}=10_{1,10}\leftarrow 9_{1,9}$ $a$-type transition of \textit{syn}-propanal is observed to be $\Delta\nu_R=295$\,kHz, if the $b$-type transition $J_{K_a,Kc}=10_{0,10}\leftarrow 9_{1,9}$ is pumped simultaneously.
Using the observed splitting and solving Eq.~\eqref{Eq2:Autler-Townes_Rabi-frequency} for $P$, an effective output power of 10\,mW within the cell is derived (a more detailed derivation of this value can be found in Eqs.~\eqref{Eq:AT_power_formula}, \eqref{Eq:AT_used_values}, and \eqref{Eq:AT_power} in the Appendix). This derived value is comparable in magnitude to the nominal output power of the pump source ($<60$\,mW), considering expected losses of windows, polarization grids, mirrors and effects of  misalignment.

\begin{figure}[t!]
\centering
\begin{tabular}[t]{cc} 
\hspace{0.3cm}
\subfloat{\includegraphics[width=0.38\linewidth]{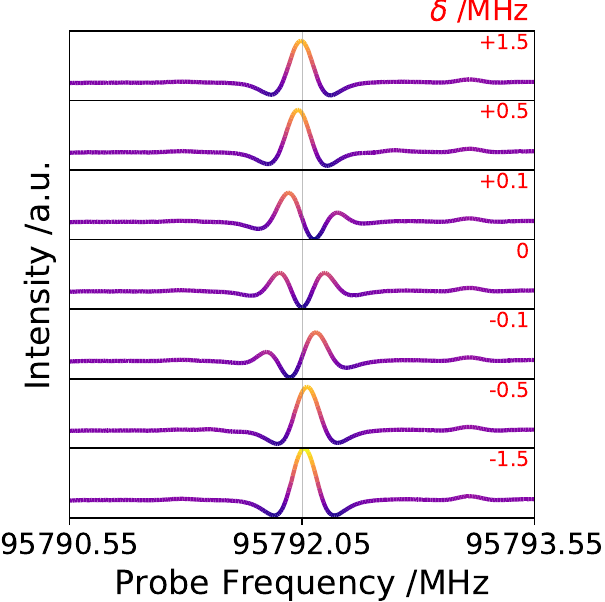}} & 
\hspace{0.3cm}
\subfloat{\includegraphics[width=0.38\linewidth]{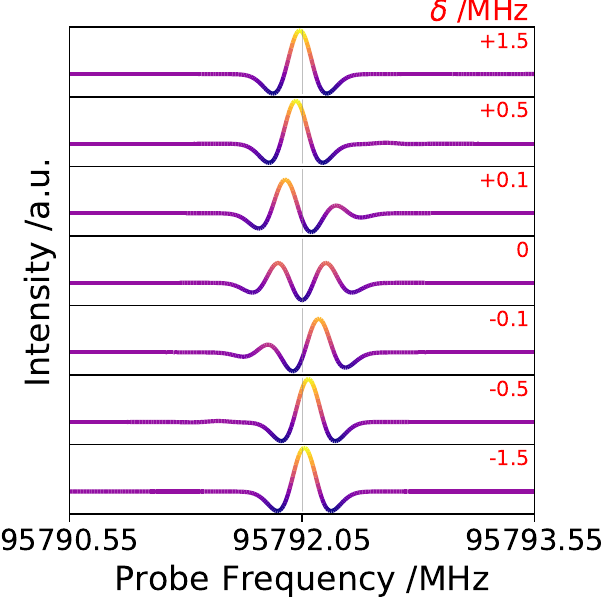}} 
\\
\subfloat{\includegraphics[width=0.41\linewidth]{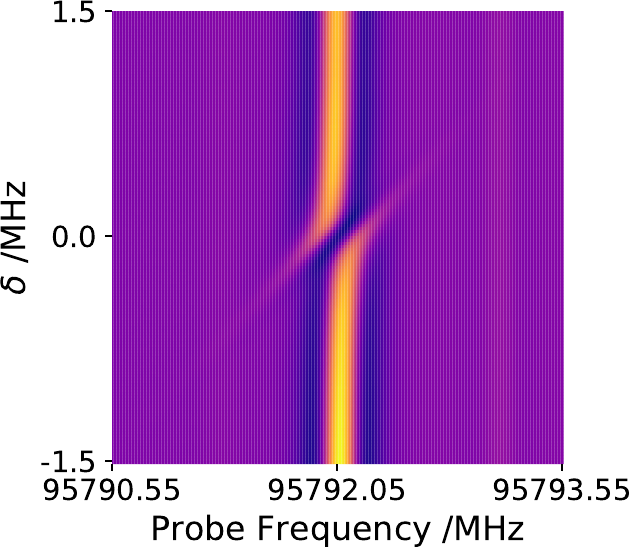}} & 
\subfloat{\includegraphics[width=0.41\linewidth]{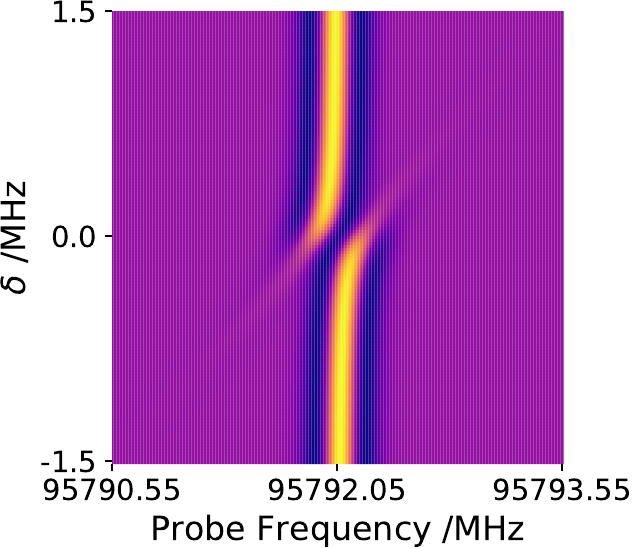}}
\\ 
\subfloat{\includegraphics[clip,trim=0 0 0 1.5cm,width=0.45\linewidth]{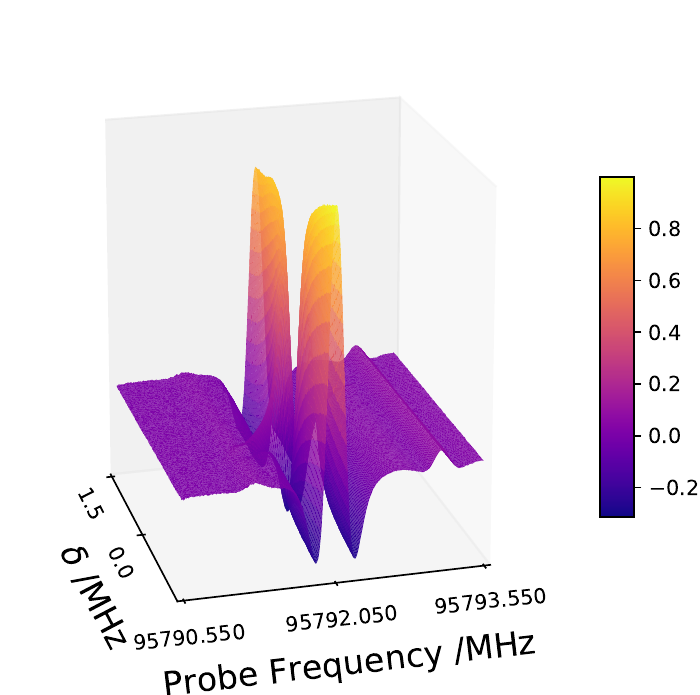}} & 
\subfloat{\includegraphics[clip,trim=0 0 0 1.5cm,width=0.45\linewidth]{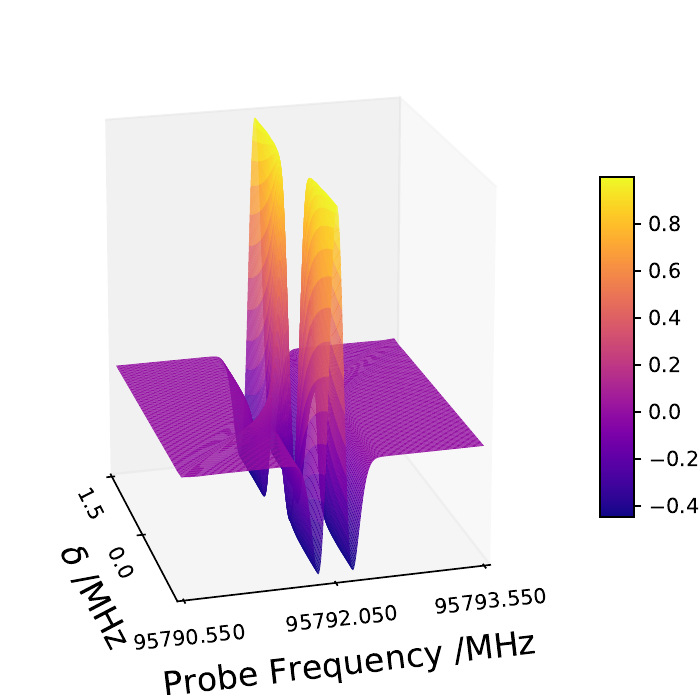}}
\end{tabular}
\caption[Autler-Townes effect in 2D DR spectra of \textit{syn}-propanal.]{
Autler-Townes effect in 2D DR spectra of \textit{syn}-propanal, shown for the same energy level arrangement as in Fig.~\ref{Fig:DR_singleLine}.
Selected single measurements (top row), a heat map (middle row), and the resulting 3D spectra (bottom row) are shown. 
On the left hand side measurements are shown and on the right hand side simulations for comparison.
More details of the measurements and simulations can be found in the text or in the Appendix, respectively.
}
\label{Fig:Autler-Townes_2D}
\end{figure}

The typical behavior of the Autler-Townes splitting (cf. Fig.~\ref{FigA:Autler_arrangements} and Eqs.~\eqref{Eq1:Autler-Townes_splitting} and \eqref{Eq3:Autler-Townes_Intensity}), in particular the asymptotic convergence of the strong component to the center frequency for larger detunings and of the steadily less intense component to a crossing diagonal, is clearly visible in a 2D DR spectrum shown in Fig.~\ref{Fig:Autler-Townes_2D}.
All 2D spectra in this work are generated by successively recording probe spectra, all with identical start-, stop-frequency, step size, and as always measured for increasing and decreasing frequencies. However, each single probe spectrum is measured with a fixed pump frequency which is increased step-wise from one to another.
In Fig.~\ref{Fig:Autler-Townes_2D}, the 3$\times$3\,MHz 2D DR spectrum is recorded with probe and pump step size of 10 and 20\,kHz, respectively. A time constant ($TC$) of 20 ms per measurement point results in a total measurement time, including computational and communication time, of roughly 77\,min.
The stronger Autler-Townes component is blue-shifted for $\delta<0$ and red-shifted for $\delta>0$, also referred to as the AC Stark shifts, as a regressive energy level arrangement is observed (cf. Fig~\ref{FigA:Autler_arrangements}). The weaker Autler-Townes component approaches a diagonal from top right to bottom left in the heat map (red-shifted for $\delta<0$ and blue-shifted for $\delta>0$).
For progressive energy level arrangements, opposite AC Stark shifts are observed and the weaker component approaches a diagonal from top left to bottom right, but somewhat smaller splittings are observed, cf. Fig~\ref{FigA:DM_arrangement_measurement}.

\subsection{DM-DR Setup - Clearing up Molecular Spectra}                 
\label{subSec:DM-DR Experiment}

Standing waves or fringes and other fluctuations in experimental conditions can be drastically reduced or even canceled by the implementation of a DM scheme, i.e. modulating both the probe and the pump frequency.
A second lock-in amplifier is used for this reason to generate the additional modulation of the pump source, cf. the optional part in Fig.~\ref{Fig:exp_setup_schematic}.
A pulse modulation (PM) of the pump source (50\,\%-on and 50\,\%-off duty cycle) with 1$f'$ demodulation subsequent to the $2f$ demodulation would be desirable intuitively.
Only probe transitions which share an energy level with the pump transition would be left in the spectrum in this case, independent of the line density of absorption spectra.
Frequency switching (FS) of the pump source is used here instead of PM to circumvent possible damages of the amplifier due to spike signals, which may occur if the power is turned on and off rapidly. FS allows to keep the pump power constant while having effectively the same effect as a PM. 
The frequency throw is performed from on-resonance ($\Delta\nu=\pm0$\,MHz) to a far away off-resonant frequency (typically $\Delta\nu=-120$\,MHz, cf. Figs.~\ref{FigA:DMDR_Detuning_Sim} and \ref{FigA:DMDR_PumpOffset}).
A far off-resonant pump frequency has a comparable effect as turning off the radiation, i.e. to the 50\,\%-off duty cycle, but without actually turning off the power.
A chopper wheel may be used as an alternative, but was observed to introduce additional noise due to vibrations and air flow to the optical components degrading the SNR of the spectrum.

\begin{figure*}[t]
\centering
\includegraphics[width=\linewidth]{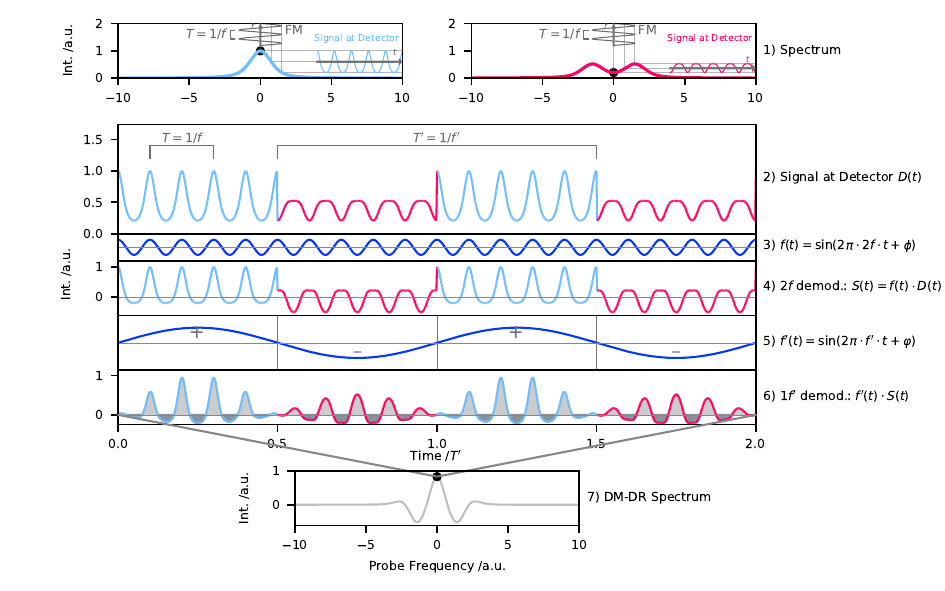}
\caption{Schematic double-modulation double-resonance (DM-DR) operation principle.
The signal at different steps of the DM-DR setup is illustrated for the determination of the intensity value of a single frequency point, here exemplarily shown on the probe center frequency.
Real line shapes for off-resonance (blue) and for on-resonance (pink) pumping are shown (row 1). The grey sine functions indicate the FM of the probe source with modulation frequency $f$. 
The signal at the detector $D(t)$ (row 2) alternates between on- and off-resonance signals due to the FS modulation of the pump source with $f'$.
By multiplying the received signal (row 2) with the 2$f$-reference signal (row 3), the 2$f$ demodulation of the first lock-in amplifier is performed (row 4).
This signal (row 4) is then multiplied with the 1$f'$-reference signal (row 5) leading to the second demodulation (row 6).
The integration of row 6, or the DC part of the electronic signal, results in one data point which is the result of the near real-time subtraction of the off- and on-resonance DR signals.
Thereby, the integration time (or time constant; $TC$) should be a multiple of $T'=1/f'$.
Repeating this procedure for several probe frequencies results in a typical baseline- and confusion-free DM-DR spectrum (row 7).
}
\label{Fig:DM_working_principle}
\end{figure*}

Two demodulation stages of the received signal are used in total for DM-DR measurements, here simply two lock-in amplifiers are connected in series for doing so, see Fig.~\ref{Fig:exp_setup_schematic}.
The applied DM-DR operation principle is shown in Fig.~\ref{Fig:DM_working_principle} and explained in detail in the following.
The spectra of a single transition, or its two blended Autler-Townes components (in blue; far off-resonant), and its split Autler-Townes components (in red; on-resonant) are depicted in the first row.
The Voigt profiles are scanned in general point by point using FM.
The first lock-in amplifier generates a 1$f$ reference signal which modulates the MMW probe frequency of the synthesizer.
The 1$f$ modulated signal passes the absorption cell and is received by the Schottky detector. The received signal is exemplarily shown in row 2 on the example of the center frequency data point. 
The pump frequency is switched from on- to off-resonance with period $T'=1/f'$ (FS), for that reason their signals depicted in blue and red are alternating likewise. Both signals are periodic to 2$f$ each as the line profile at the center frequencies are symmetric.
The 2$f$ demodulation, multiplication of the detector signal with a sine wave with frequency 2$f$ (row 3), leads to the 2$f$ demodulated signal in row 4.
An integration of either the blue or the red part, without taking into account the 1$f'$ alternation, would lead to the center frequency data point of the $2f$ conventional (Sec.~\ref{subSec:Convenional Experiment}) or DR (Sec.~\ref{subSec:DR Experiment}) spectra (cf. line profiles in Fig.~\ref{Fig:DR_singleLine}), respectively.
However, the $2f$ demodulated signal is transferred additionally to a second lock-in amplifier for DM-DR measurements.
The additional $1f'$ demodulation is experimentally realizing the difference spectrum of the on- and off-resonant measurements (or, in other words, a subtraction of the conventional and the DR measurement).
A descriptive illustration of the subtraction procedure is visualized by taking a look at the period $T'=1/f'$ in rows 4 and 5.
The subtraction is performed as the sign of the sine wave of the $1f'$ reference signal (row 5) is positive from $0$ to $\Pi$ (or from $nT'$ to $(2n+1)T'/2$ with $n\in\mathbb{N}$), in particular when the pump radiation is off-resonance (blue in row 4), and it is negative from $\Pi$ to $2\Pi$ (or from $(2n+1)T'/2$ to $nT'$), in particular when the pump radiation is on-resonance (red in row 4).
Finally, the integration over a time period $TC$ of the signal, or the DC part of the electronic signal, shown in row 6, being the multiplication of rows 4 and 5, results in one data point of a DM-DR spectrum (row 7).
It is imperative that the reference frequency of the first lock-in amplifier $f$ is higher than $f'$ of the second lock-in amplifier.
Ideally $f$ is a multiple of $f'$ and $TC$ is a multiple of $T'=1/f'$, such that always exactly the same number of on- and off-resonant cycles are averaged. A reference frequency $f'\sim 200$\,Hz was found to be favorable and used for all subsequent measurements, cf. Fig.~\ref{FigA:DMDR_Modulationsfrequenz}. The most important measurement parameters are collected in Table~\ref{Tab:Parameters}.
This procedure is repeated for several probe frequencies to recreate the DM-DR line profile shown in row 7. The resulting line profile is accidentally similar to a conventional 2$f$ measurement, striking are, however, two small additional local maxima which occur as the two Autler-Townes components being "pushed" away from the unperturbed, or conventional, center frequency.

\begin{table}[t]   
\caption[]{Typical measurement parameters of DM-DR measurements.}
\setlength{\tabcolsep}{0pt}
\begin{tabular}{llrl}
Parameter & Device & \multicolumn{2}{l}{Value} \\
\toprule
$f$ & 1$^{st}$ lock-in amplifier & $\approx~~47$&\,kHz \\
FM amplitude~~~~~ & probe synthesizer & \multicolumn{2}{l}{$\approx FWHM$}  \\
$f'$ & 2$^{nd}$ lock-in amplifier~~~~~~ & $\approx200$&\,Hz\\
FS throw & pump synthesizer & $-$120 &\,MHz \\
$TC$ & 2$^{nd}$ lock-in amplifier & 50&\,ms \\
\bottomrule
Ideally:\\
\multicolumn{4}{l}{ $mT=m/f = nT'=n/f'= TC $ with $m,n\in\mathbb{N};~f>f'$}
\label{Tab:Parameters}     
\end{tabular}
\end{table}

In Figs.~\ref{Fig:DR_singleLine} and \ref{Fig:Autler-Townes_2D}, the observed splitting is ideal for maximizing the DM-DR signal, therefore, the effective output pump power within the cell is optimal $P=P_{opt}$ with respect to the observed 3-level system.
The best possible DM-DR signal is obtained if the higher frequency minimum of the red-shifted 2$f$ Autler-Townes component and the lower one of the blue-shifted component of the DR measurement are coinciding at the center frequency of the conventional measurement.
More details can be found in Figs.~\ref{FigA:DMDR_PumpPowers_Sim} and \ref{FigA:DMDR_Luis_Max}.

\section{Experimental Results}                
\label{Sec:ExperimentalResults}               

Two main advantages of taking MMW-MMW DM-DR spectra will be demonstrated in this section. First, identical to DR measurements, linkages can be easily determined (Sec.~\ref{SubSec:Exp_Linkages}), but more importantly with DM-DR measurements  also weak and blended absorption features can be identified easily (Sec.~\ref{SubSec:Exp_weakLines}).

\subsection{Finding Linkages}                 
\label{SubSec:Exp_Linkages}

\begin{figure}[t]
\centering
\includegraphics[width=\linewidth]{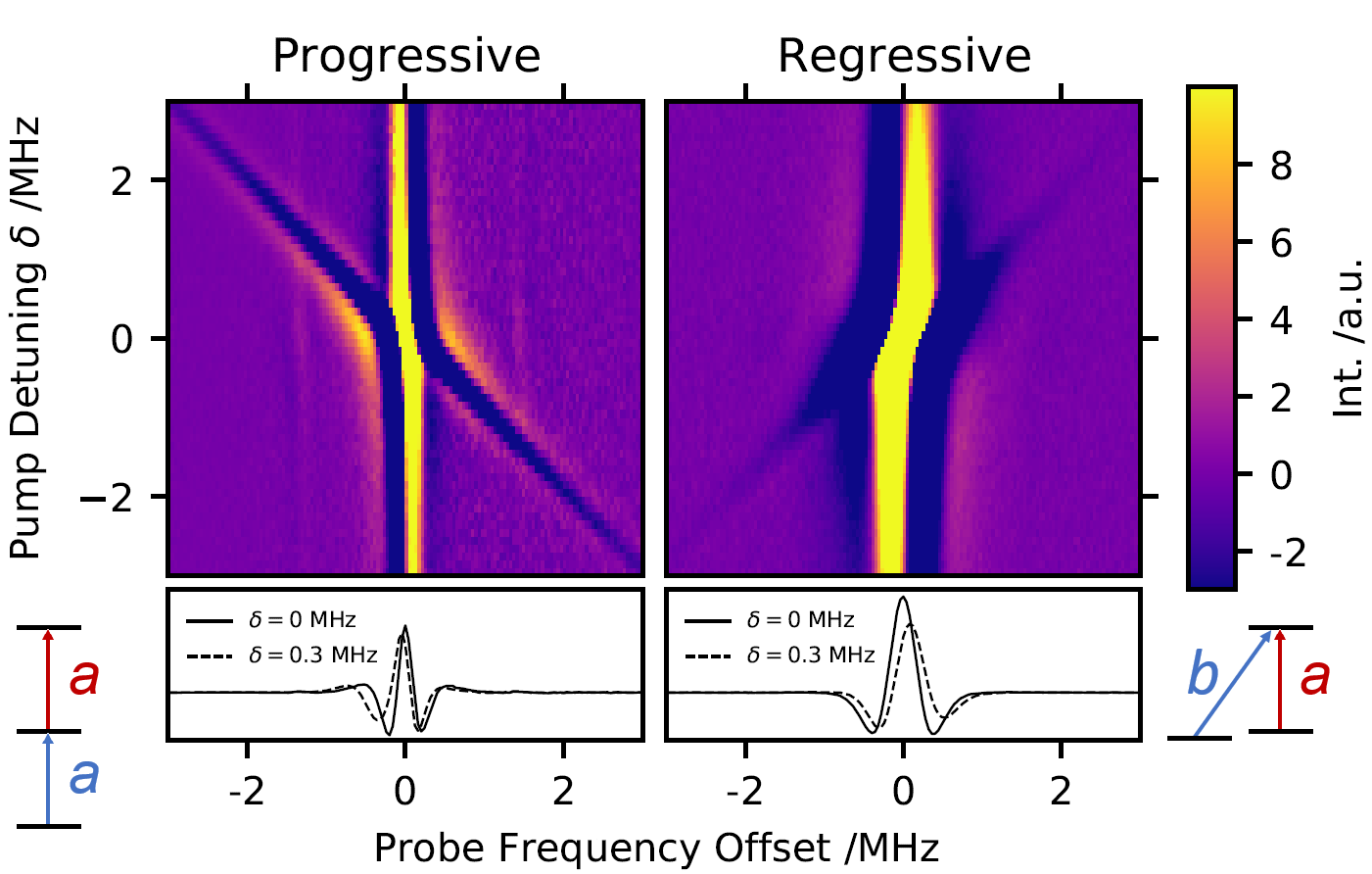}
\caption{
Two heat maps of 2D DM-DR experimental spectra ($6\times6$\,MHz) are shown for ethyl cyanide for a progressive (left) and for a regressive energy level arrangement (right).  
Progressive and regressive energy ladder arrangements can be easily distinguished by red- or blue-shifted Autler-Townes components. For more details see text.
}
\label{Fig:ProReg}
\end{figure}

Linked transitions, which share an energy level, can easily be found with DR spectroscopy. A rather straightforward, albeit time consuming, method of finding linkages is to record 2D spectra, e.g. 2D MMW-MMW DM-DR spectra (Fig.~\ref{Fig:ProReg}).
2D DM-DR experimental spectra with dimensions of $6\times6$\,MHz are shown for ethyl cyanide, both for a progressive and a regressive energy level arrangement.
The $J_{K_a,K_c}=12_{1,12}\leftarrow 11_{1,11}$ $a$-type transition is used for both 3-level systems as pump transition (104051.276\,MHz). The $J_{K_a,K_c}=11_{1,11}\leftarrow 10_{1,10}$ $a$-type transition (95442.478\,MHz) is probed to study a progressive arrangement, whereas the $J_{K_a,K_c}=12_{1,12}\leftarrow 11_{0,11}$ $b$-type transition (115341.909\,MHz) is probed for a regressive one. 
Both spectra are measured with 40\,kHz probe steps and 100\,kHz pump steps.
The $TC$s are chosen to be $TC=50$\,ms (total 21\,min) and 5$\times$100\,ms (173\,min), respectively, as the dipole moment component of the two probe transitions $\mu_a=3.816$\,D and $\mu_b=1.235$\,D are rather different \cite{KRASNICKI201183}, hence SNRs are comparable even though in both 2D spectra the SNRs are so high that in principle the total measurement time could be reduced.
Again, progressive and regressive energy ladder arrangements can be distinguished by opposing AC Stark shifts. It should be noted that the DM character basically performs a subtraction of "conventional$-$DR" and therefore the AC Stark shift seems to be opposite in DM-DR measurements compared to DR ones. The splitting of regressive energy ladder arrangements is somewhat larger as can be seen from the broader line profile (broader yellow feature in Fig.~\ref{Fig:ProReg}).
The diagonal crossing, originating in the hyperbolic behavior of the weaker Autler Townes component, from top left to bottom right or from top right to bottom left is visible for progressive and regressive energy level arrangements, as in DR measurements, respectively, but its magnitude is inverted in sign.
A $30\times30$\,MHz 2D DM-DR spectrum of ethyl cyanide linking more than two transitions with each other is shown in the Appendix in addition to demonstrate that even multiple systems, close in frequency for both pump and probe transitions, can be disentangled (Fig.~\ref{Fig:EtCN_2D}).

\subsection{Identifying Weak Features}         
\label{SubSec:Exp_weakLines}                  

\begin{figure}[t]
\centering
\includegraphics[width=\linewidth]{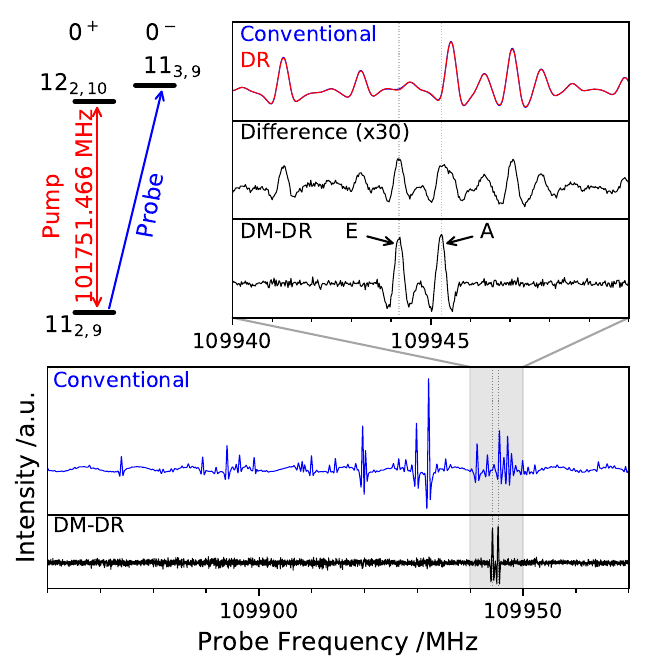}
\caption[Conventional vs DR vs DM-DR measurements.]
{
The baseline- and confusion-free character of DM-DR measurements is highlighted.
In the zoomed portion of the spectrum, a comparison of a conventional measurement overlaid by a DR measurement (top row), their calculated difference spectrum (conventional$-$DR; middle row), and a DM-DR spectrum (bottom row) of \textit{gauche}-propanal are shown. The observed three level system is shown on the top left.
DM-DR measurements show two lines as the probed $c$-type transition, which connect the two tunneling states $0^+$ and $0^-$ of the vibrational ground state, shows internal rotation splitting (A and E components), but both components are blended at the pump frequency. Both components are heavily blended and thus not assignable in conventional measurements.  
}
\label{Fig:Propanal_weak_ctype}
\end{figure}

Weak transitions are frequently blended by stronger features due to the high density of transitions of COMs in the (sub-)MMW region and, therefore, proper detection or assignments may not be possible. 
Nevertheless, these transitions may contain valuable spectroscopic information that may be instrumental for deriving important parameters of molecular Hamiltonians.
Weak $c$-type transitions of \textit{gauche}-propanal, which were not assigned in an earlier study \cite{ZINGSHEIM2017_Propanal}, may serve as an illustrative example.
Thanks to the DM character of the presented setup, even these weak and blended target features can now be identified with ease (Fig.~\ref{Fig:Propanal_weak_ctype}).
The importance of such transitions for a proper analysis of the \textit{gauche}-propanal spectrum will be described elsewhere.

Simple subtraction of a conventional and a DR measurement, which are taken one after another, may result in false positive signals, if the experimental conditions are not sufficiently stable and vary between measurements (cf. middle row of the zoom in Fig.~\ref{Fig:Propanal_weak_ctype}). In the worst case, target lines are still blended.
Fortunately, the DM-DR technique is only sensitive to fluctuations of the experimental setup/conditions on a time scale smaller than the $TC$ (tens of milliseconds), in particular not to slight pressure changes in the gas cell, hence, false positives are much less an issue (cf. bottom row in Fig.~\ref{Fig:Propanal_weak_ctype}).
Furthermore, the experimental subtraction leads to baseline-free spectra, as standing waves mainly originate from the probe radiation (as the polarized pump beam is not reaching the detector).
Weak and blended transitions can hence unambiguously be assigned in a straightforward fashion, as can be seen in the lower panel of Fig.~\ref{Fig:Propanal_weak_ctype}.

\section{Discussion and Conclusions}           
\label{Sec:Discussion}

The Autler-Townes effect makes DR and DM-DR measurements a useful tool for rotational spectroscopy in the MMW-MMW region.
The DM character generally results in baseline- and confusion-free spectra.
This is especially useful for the very dense spectra observed for COMs and has been demonstrated here on the selected examples ethyl cyanide and propanal.
The big advantage of MMW-MMW DM-DR measurements is the unambiguous assignment of very weak or blended transitions (Fig.~\ref{Fig:Propanal_weak_ctype}).

The best possible DM-DR signal is obtained if the left minimum of the blue shifted component and the right minimum of the red shifted component of the 2$f$ line shape are blended at the unperturbed center frequency, which leads to the largest possible difference of conventional and DR measurements, cf. Figs.~\ref{FigA:DMDR_PumpPowers_Sim} and \ref{FigA:DMDR_Luis_Max}.
This holds for the measured pair of transitions of propanal in Fig.~\ref{Fig:DR_singleLine}.
In this case, in combination with the given dipole moment of the pumped transition, the effective output power within the cell is "by accident" optimal, $P=P_{opt}$, in the sense that the DM-DR signal would be maximized.

Sometimes it may occur that both pump and probe frequencies of more than one three level system are close to each other. In this case, additional signals may appear in the DM-DR signal (bottom row of Fig. \ref{Fig:EtCN_2D}). However, these signals can be identified by their accidental 1$f$-line shape because pump frequencies are slightly off-resonance (cf. also Fig.~\ref{FigA:DMDR_Detuning_Sim}). 
Furthermore, an imperfectly aligned pump radiation may be received by the Schottky detector which can lead to small standing waves (imperfect polarization).
If the off-resonant frequency of the frequency throw is connecting another third energy level to the probed two-level system by chance, it will cause an inverted line shape of the DM-DR signal as it can be interpreted as a phase shift of the demodulated signal of 180\,$^\circ$.
It is important to ensure that the off-resonant frequency throw of the FS procedure is sufficiently large and that the pump frequency is determined exactly, otherwise the so-called AC Stark shift (off-resonant pumping of the Autler-Townes effect), results in DM-DR line shapes where the maximum intensity is not coinciding with the center frequency, cf. Fig.~\ref{FigA:DMDR_PumpOffset}.

The SNRs of DM-DR measurements with regressive arrangements are observed to be larger than those of progressive ones, which arises from larger observed splittings, cf. Figs.~\ref{Fig:ProReg} and \ref{FigA:DM_arrangement_measurement}.
The origin of this finding is planned to be studied in more detail in the future.
The difference is expected to originate from small splittings due to internal rotation, which are, however, blended in the conventional spectra of pump and probe transitions. This unresolved splitting would lead to very small off-resonance pumping of A and E components.
For regressive arrangements, the internal rotation component lower in frequency is expected to shift to even lower frequencies (as the pump is blue shifted for this component; $\delta>0$), whereas the higher one to even higher frequencies ($\delta<0$), which results effectively in larger splittings than opposite shifts for progressive arrangements. In fact, the true Autler-Townes splitting would therefore be somewhat smaller than the determined one and therefore also the power in the cell is somewhat smaller than 10\,mW. Molecules without any spectroscopic substructure, originating from, e.g., internal rotation, or hyperfine structure should be ideal targets to study the impact of different energy level arrangements. However, energy term diagrams, or the resulting spectra, of COMs frequently feature such spectroscopic substructure.

\section{Outlook}           
\label{Sec:Outlook}

This work demonstrates that 2D DM-DR spectroscopy is a very powerful technique to disentangle weak signals in dense spectra of many molecules.
Nevertheless, it would be beneficial to further speed up the experimental procedure.
It may be sufficient to link only center frequencies of lines with each other, instead of recording full 2D spectra, i.e. to monitor the intensity modulation of a single frequency, to identify linkages.
An accurate line list of the molecular fingerprint is required to do this. The major challenge is probably to set a proper threshold which discriminates real linkages from AC Stark shifted signals, which may originate from other 3-level systems that are close in pump and probe frequencies and which can be identified by their line shape in 2D spectra (cf. Fig.~\ref{Fig:EtCN_2D}).
Of course, 2D measurements can be sped up with already known fast scanning experimental techniques
\cite{Petkie-DeLucia1997_FASSST,AlekseevMotiyenkoMargules_2012,Hays2016_FastSweep,Zou2020_EmissionDDS,Balle_Flygare_1981,Brown_Pate_2008}.

The DM character can be extended to many other experimental setups.
In CP experiments for example, spectra with an additional radiation source (DR on), which is able to destroy the phase coherence of the observed free-induction decay, may be taken with a CP shifted in phase by 180$^\circ$ out-of-phase relative to the DR off measurement.
In this way, a comparable difference spectrum as for the DM-DR experimental setup performed in this work is created when both spectra are added up (cf., e.g., Ref.~\cite{Pate2010_Coherence-converted}).
In general, all kinds of additional sources
may be modulated as well, as already mentioned in the introduction and demonstrated in the literature \cite{AMANO1974,AMANO2000_DM_HOC,Gudeman_Saykally_1983,Havenith1994_DM_concentration_Ar-CO}. In fact, it does not matter whether the additional modulation (or intensity manipulation) is based on the Autler-Townes splitting, population difference, Zeeman splitting or any other mechanism. DR can be sensitive to single target transitions, whereas, for example, a modulation of a magnetic field only separates between transitions from closed and open-shell molecules.

Given the splitting of 295\,kHz of the system presented in Fig.~\ref{Fig:DR_singleLine}, intensity changes of about 10\,\%, based on the Autler-Townes splitting, are still expected for comparable COMs with dipole moments as low as about 0.3\,D with the current experimental setup (see Appendix, Eq.~\eqref{Eq:Dipole_for_0.1}, for a more detailed derivation of this value).
Higher output powers are needed for maximizing the DM-DR signals from smaller dipole moments.
In addition, also population changes due to Rabi oscillations alter the intensity considerably. These are neglected in the discussion of the results presented here as the Autler-Townes effect is dominating the intensity change. This becomes particularly important when only smaller output powers are available due to technical reason, for example at smaller wavelengths. 
The DM-DR technique is not at all limited to MMW-MMW spectroscopy.
In principle, pump and probe DM-DR experiments in different frequency regions are only limited by existing linkages (allowed transitions between 3-level systems.)

One major idea is to automate the analysis of rotational fingerprints of COMs in the future.
After successfully linking two candidate transitions, the connection of more lines can be tested in the fashion of the AUTOFIT routine \cite{SEIFERT2015_Autofit}, where from assumed assignments, next candidate lines are found. The benefit of DM-DR measurements is that linkages can be proven experimentally with ease and wrong "solutions" can be discarded without any doubt, contrary to the search of the statistically most relevant solutions.
Systematic procedures of using DR linkages are already known in the literature, see e.g. Refs.~\cite{Crabtree2016_Taxonomy,Martin-Drumel2016_AMDOR}.
The main idea is to merge smart algorithms which use ab-initio calculations, then verify assignments experimentally, eventually deriving proper quantum mechanical models in an automated iterative fashion of linking (and with that assigning), fitting, and modeling.

The DM-DR experimental setup is expected to simplify the analysis of strongly perturbed spectra and the unambiguous assignment of weak and blended transitions may support this.
In general, greatly simplified spectra on the basis of baseline- and confusion-free DM-DR measurements can facilitate the deciphering of complex molecular fingerprints.

\section*{Acknowledgement}
This work has been supported via Collaborative Research Centre 956, sub-projects B3, funded by the Deutsche Forschungsgemeinschaft  (DFG; project ID 184018867) and DFG SCHL 341/15-1 
(``Cologne Center for Terahertz Spectroscopy").

\section*{Appendix: Supplementary data}

\setcounter{figure}{0}
\renewcommand{\thefigure}{A\arabic{figure}}
\setcounter{equation}{0}
\renewcommand{\theequation}{A\arabic{equation}}

At first, the derived observable splitting or the Rabi frequency $\nu_R$ for on-resonant pumping, in Eq.~\eqref{Eq2:Autler-Townes_Rabi-frequency}, is elucidated in more detail than in the main paper. 
The formula for $\nu_R$ for on-resonance pump frequencies ($\delta=0$), can be determined with the help of $\Delta E =E_1-E_2= \hbar \omega_R$ (from Eq.~\eqref{Eq1:Autler-Townes_splitting} \cite{AutlerTownes1955,Cohen-Tannoudji2008_DressedAtomApproach}) and the following relations, which can be found for example in Ref.~\cite{Bernath}:
\begin{subequations}
\begin{align}
I&=P/A \\
A&=\pi d^2/4  \\
I&=\frac{1}{2}\epsilon_0cE^2 \\
\hbar\omega_R&=h\nu_R \\
\hbar\omega_R&=\mu_{ab}E
\end{align}
\label{Eq:AT_power_formula}
\end{subequations}
with the intensity $I$ of the pump beam, its area $A$, its output power $P$, its diameter $d$, its field intensity $E$, the transition dipole moment of the 2-level system $\mu_{ab}$ with Rabi frequency $\omega_R$, the vacuum permittivity $\epsilon_0$ and the speed of light $c$.

From Eq.~\eqref{Eq2:Autler-Townes_Rabi-frequency},
the effective output power of the pump source $P$, can be determined by using the following values of the observed system from Fig.~\ref{Fig:DR_singleLine}:
\begin{equation}
\nu_R=295\,\text{kHz},~\mu_b=1.85\,\text{D~\cite{doi:10.1063/1.1725377}},~\text{and}~d=10\,\text{cm}.
\label{Eq:AT_used_values}
\end{equation}
The effective output power of the pump source $P$ within the cell is
\begin{equation}
\Rightarrow P\approx 10\,\text{mW}.
\label{Eq:AT_power}
\end{equation}

\begin{figure}[t]
\centering
\includegraphics[width=0.9\linewidth]{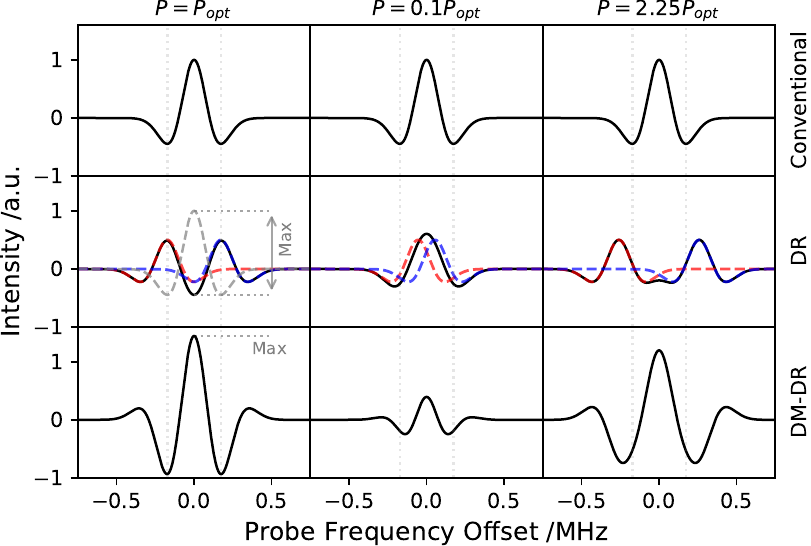}
\caption{
Simulated DM-DR line shapes for different output powers of the pump source $P$ in relation to the optimal power $P_{opt}$, which maximizes the DM-DR signal, cf. Fig.~\ref{FigA:DMDR_Luis_Max}.
The DM-DR signal (bottom row) is basically the difference of a conventional (top row) and a DR measurement (middle row), cf. Fig.~\ref{Fig:DM_working_principle}.
The pump source is off for conventional measurements (DR off: $P=0$\,W).
The two Autler-Townes components and their sum are shown in blue, red, and black for DR measurements, respectively. 
The maximum signal at DM-DR measurements is obtained if the left minimum of the blue shifted component and the right minimum of the red shifted component are blended at the unperturbed center frequency ($P$=$P_{opt}$; $\omega_R\propto \sqrt{P}$ \cite{Bernath}).
}
\label{FigA:DMDR_PumpPowers_Sim}
\end{figure}

\begin{figure}[t]
\centering
\includegraphics[width=0.9\linewidth]{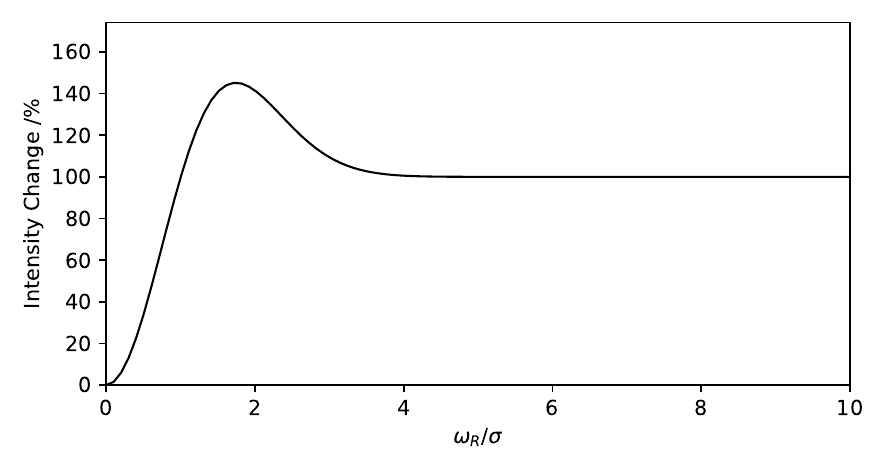}
\caption{
The intensity change between DR off and on measurements of the center frequency results in the signal amplitude of the DM-DR measurement.
The simulated change for 2$f$ components is plotted here as variable of the Rabi frequency $\omega_R$ (depending on the pump power $P$ and the transition dipole moment $\mu_{ab}$, see Eq.~\eqref{Eq:Rabi_dip+E}), which determines the magnitude of the splitting of the Autler-Townes components, and the standard deviation of the Gaussian $\sigma$ (or line width). 
The exact curvature depends on the FM amplitude. Nevertheless, a maximum is observed for an optimal output power $P_{opt}$, see Fig.~\ref{FigA:DMDR_PumpPowers_Sim}.
}
\label{FigA:DMDR_Luis_Max}
\end{figure}

\begin{figure}[t]
\centering   
\includegraphics[width=0.9\linewidth]{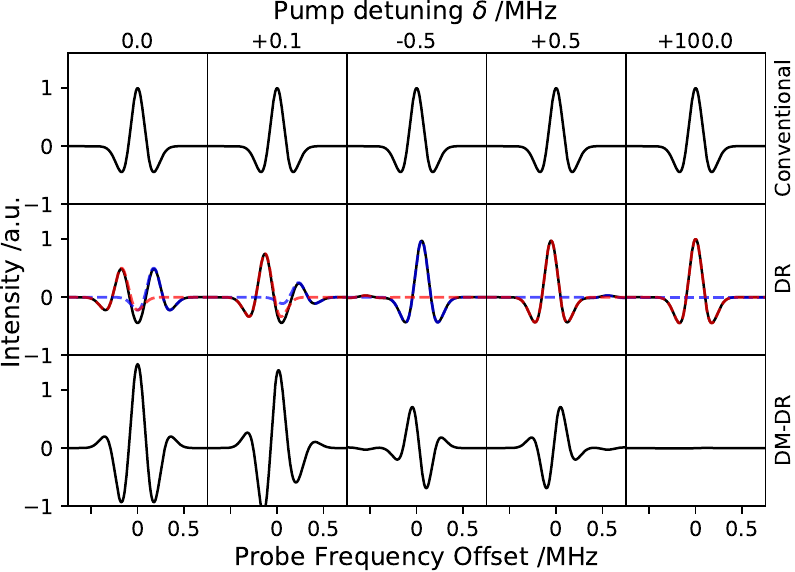}
\caption{
Simulated DM-DR line shapes for different detunings of the pump frequency $\delta$ of a regressive energy level arrangement.
The DM-DR signal (bottom row) is basically the difference of a conventional (top row) and a DR measurement (middle row). 
The output power is the same for all DR and DM-DR spectra and is chosen to maximize the DM-DR signal for zero detuning ($P=P_{opt}$, cf.~Figs.~\ref{FigA:DMDR_PumpPowers_Sim} and \ref{FigA:DMDR_Luis_Max}).
The two Autler-Townes components and their sum are shown in blue, red, and black for DR measurements (middle row), respectively.
The DM-DR line shapes are accidentally close to a first derivative of a Gaussian, or close to the 1$f$ line shape of a conventional measurement, for detuned pump frequencies, here $\delta=\pm0.5$\,MHz.
Slightly off-resonant pumping is noticed by asymmetric line shapes for DR and DM-DR measurements, see $\delta=+0.1$\,MHz. 
On the other hand, DM-DR line shapes are accidentally close to a second derivative of a Gaussian, or close to the 2$f$ line shape of a conventional measurement, for on-resonant pumping. 
Far off-resonant pumping, here $\delta=100$\,MHz, results in basically no intensity change and therefore no measurable DM-DR signal.
}
\label{FigA:DMDR_Detuning_Sim}
\end{figure}

\begin{figure}[t]
\centering
\includegraphics[width=\linewidth]{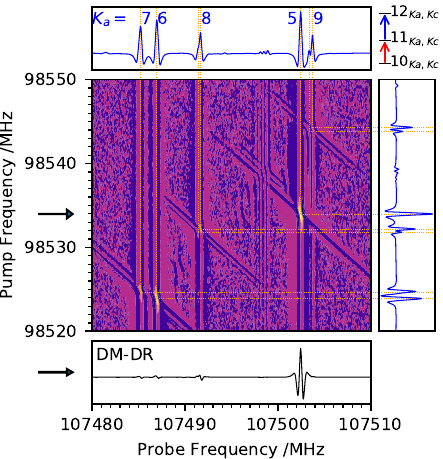}
\caption{
A heat map of a 2D DM-DR experimental spectrum is shown for ethyl cyanide ($30\times30$\,MHz, 50\,kHz probe steps and 200\,kHz pump steps with $TC$ $2\times50$\,ms results in 6.4\,h total data acquisition time). The linkages for five different combinations of pump and probe transitions with $K_a=5-9$, for the progressive energy ladder arrangement shown on the top right, are clearly visible. All transitions are prolate paired and the internal rotation components A and E, originating from a large amplitude motion of the $-$CH$_3$ group, are blended, except for $K_a=8,9$ where two lines are partly resolved.
Two conventional spectra (top and right) are shown to illustrate the linkages.
The DM-DR spectrum shows one single measurement of the 2D spectrum, pointed out by the arrow. Signals at the other linked probe frequencies are visible but can be distinguished by their magnitudes and their line shapes which look accidentally like a 1$f$ feature (cf. Fig~\ref{FigA:DMDR_Detuning_Sim}). Closer inspection further reveals a linked triplet around 107498.5\,MHz and 98539\,MHz.
}
\label{Fig:EtCN_2D}
\end{figure}

Furthermore, additional simulations for DM-DR measurements with 2$f$ line shapes are given (Figs.~\ref{FigA:DMDR_PumpPowers_Sim}$-$\ref{FigA:DMDR_Detuning_Sim}).
First, expected DM-DR line shapes in the W-band region for three different pump powers are shown to highlight the dependence of the splitting of the Autler-Townes components on the power ($\omega_R\propto \sqrt{P}$ \cite{Bernath}, see Fig.~\ref{FigA:DMDR_PumpPowers_Sim}).
This is done to highlight the need for high enough output powers of the pump source.
The best DM-DR signal is obtained if the left minimum of the blue shifted component and the right minimum of the red shifted component are blended at the unperturbed center frequency, in this case $P=P_{opt}$.
The dependence of the maximum amplitude of the DM-DR signal, or the intensity change between DR off and on measurements of the center frequency, is shown in Fig.~\ref{FigA:DMDR_Luis_Max}.
Considering the intensity change in Fig.~\ref{Fig:DR_singleLine} to be maximized (145\,\% with $\mu_{ab}=1.85$\,D), the corresponding $\omega_R/\sigma$ is determined to $\approx 1.74$ (this value can be directly read off from Fig.~\ref{FigA:DMDR_Luis_Max}).
An intensity change of about 10\% accounts for $\omega_R/\sigma\approx0.26$.
Under identical experimental conditions, in particular identical output powers and identical line widths it follows from the expression $\omega_R/\sigma=\mu_{ab}E/(h\sigma)\propto \mu_{ab}$ that $\omega_R/\sigma$ linearly depends on $\mu_{ab}$ as the other parameters are constant. Therefore, the required dipole moment component, resulting in an intensity change of 10\,\% can be calculated as 
\begin{equation}
\mu_{ab}^{0.1}=\frac{\omega_R^{0.1}/\sigma}{\omega_R^{max}/\sigma}\mu_{ab}^{max}\approx \frac{0.26}{1.74}\cdot 1.85\,D\approx 0.3\,D
\label{Eq:Dipole_for_0.1}
\end{equation}
Note that this estimation also depends on the applied FM amplitude, which has an influence on the curvature in Fig.~\ref{FigA:DMDR_Luis_Max}. 
For further clarification of expected DM-DR signals, also the line shapes for different off-resonant detunings are shown (Fig.~\ref{FigA:DMDR_Detuning_Sim}).

\begin{figure*}[t]
\centering
\includegraphics[width=0.44\linewidth]{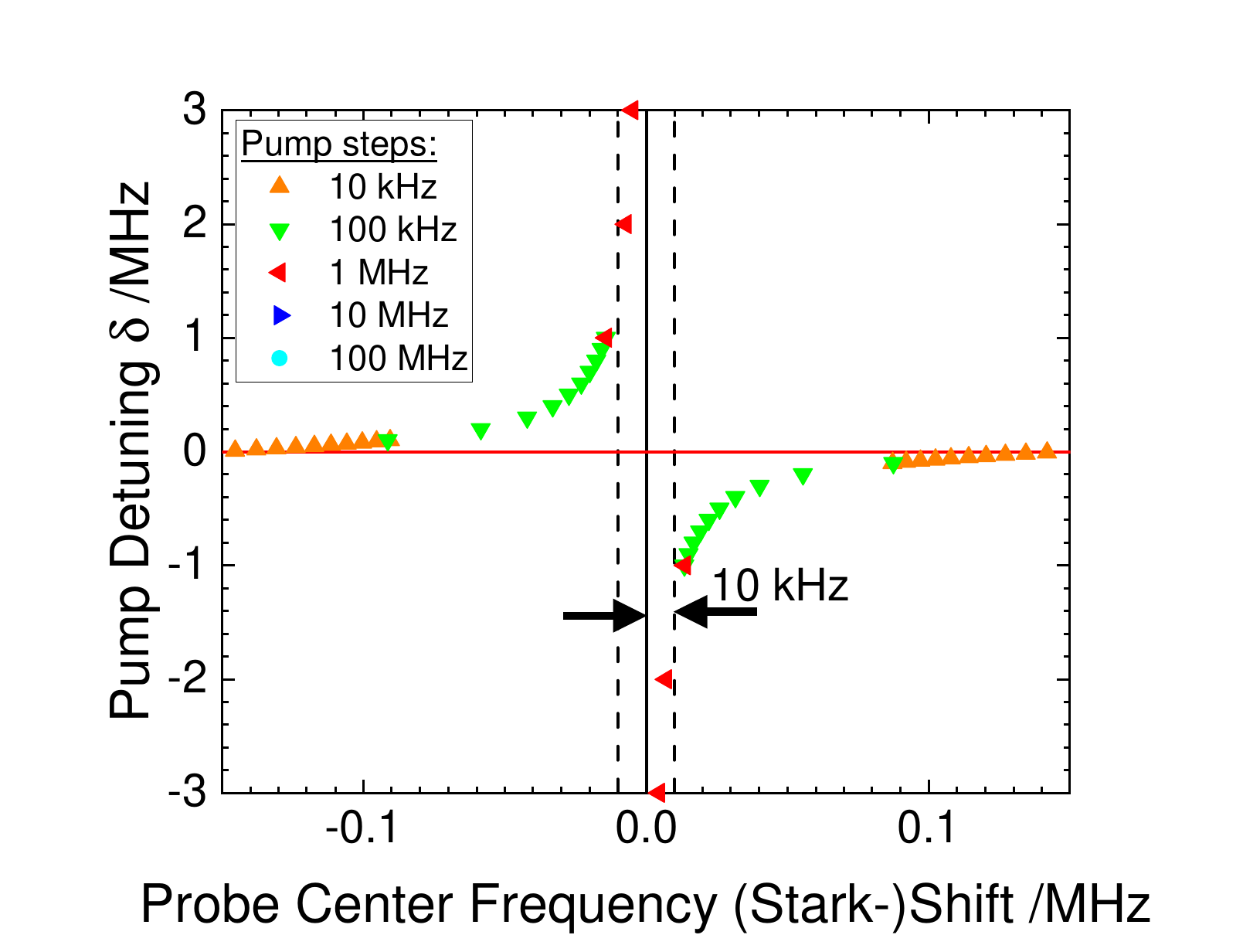}
\includegraphics[width=0.44\linewidth]{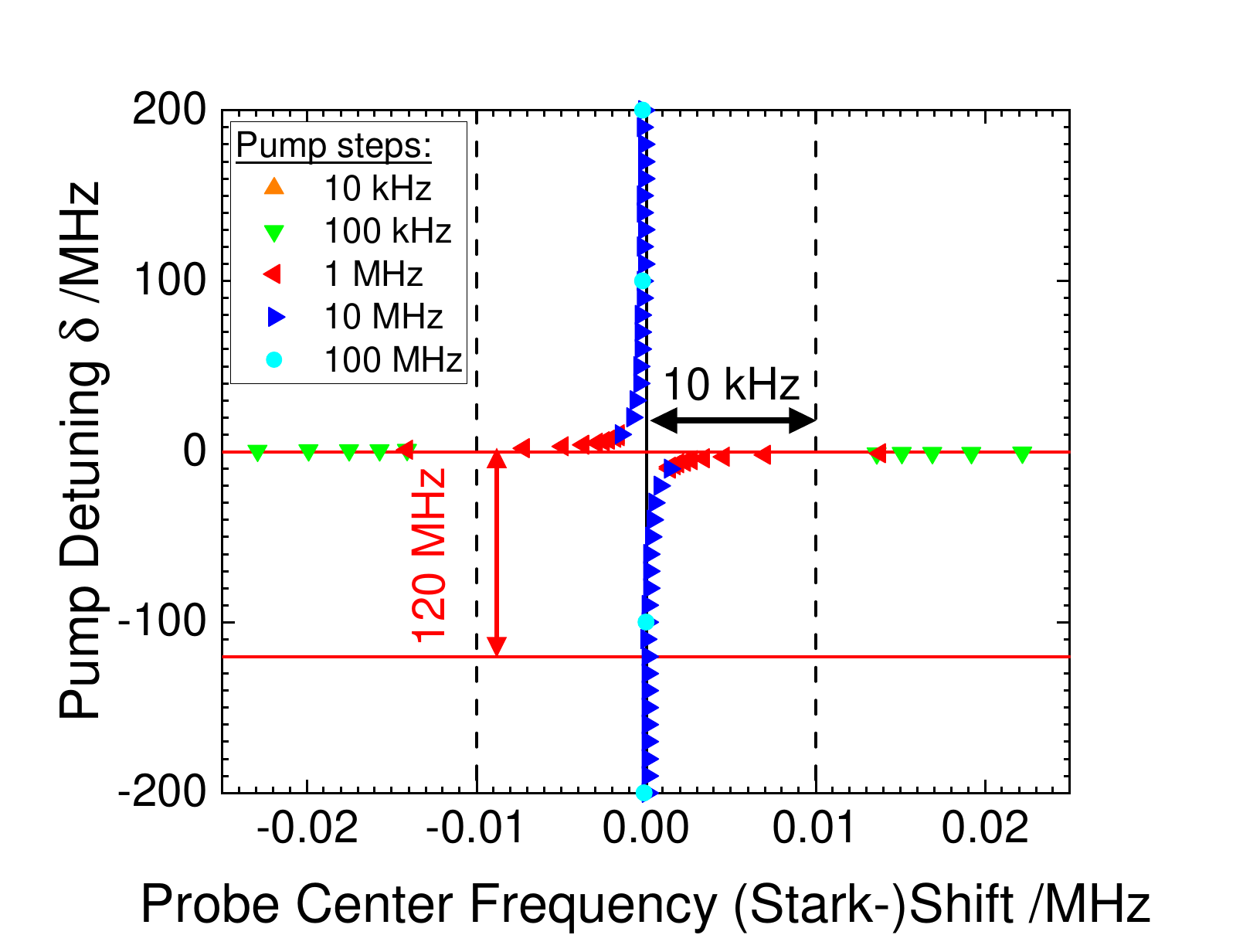}
\caption{
Determined probe center frequencies of the stronger Autler Townes component (x-axis), often also referred to as the AC Stark shift, dependent on the detuned pump frequency (y-axis) for a conventional DR measurement.
The pump-probe system is the same as in Fig.~\ref{Fig:DR_singleLine}.
The different colors and markers have no physical relevance and are used to clarify further the detuning of the pump frequency and both figures show the same measurements but with different axes.
Center frequencies of strong and symmetric lines are usually given with an uncertainty of 10\,kHz, cf. dashed lines.
For DM-DR measurements, the off-resonant DR frequency is shifted by 120\,MHz compared to the on-resonant one. A 120\,MHz off-resonant pump frequency results in negligible AC Stark shifts and has therefore a similar effect as a completely turned off pump power.
}
\label{FigA:DMDR_PumpOffset}
\end{figure*}

\begin{figure*}[t]
\centering
\includegraphics[width=0.99\linewidth]{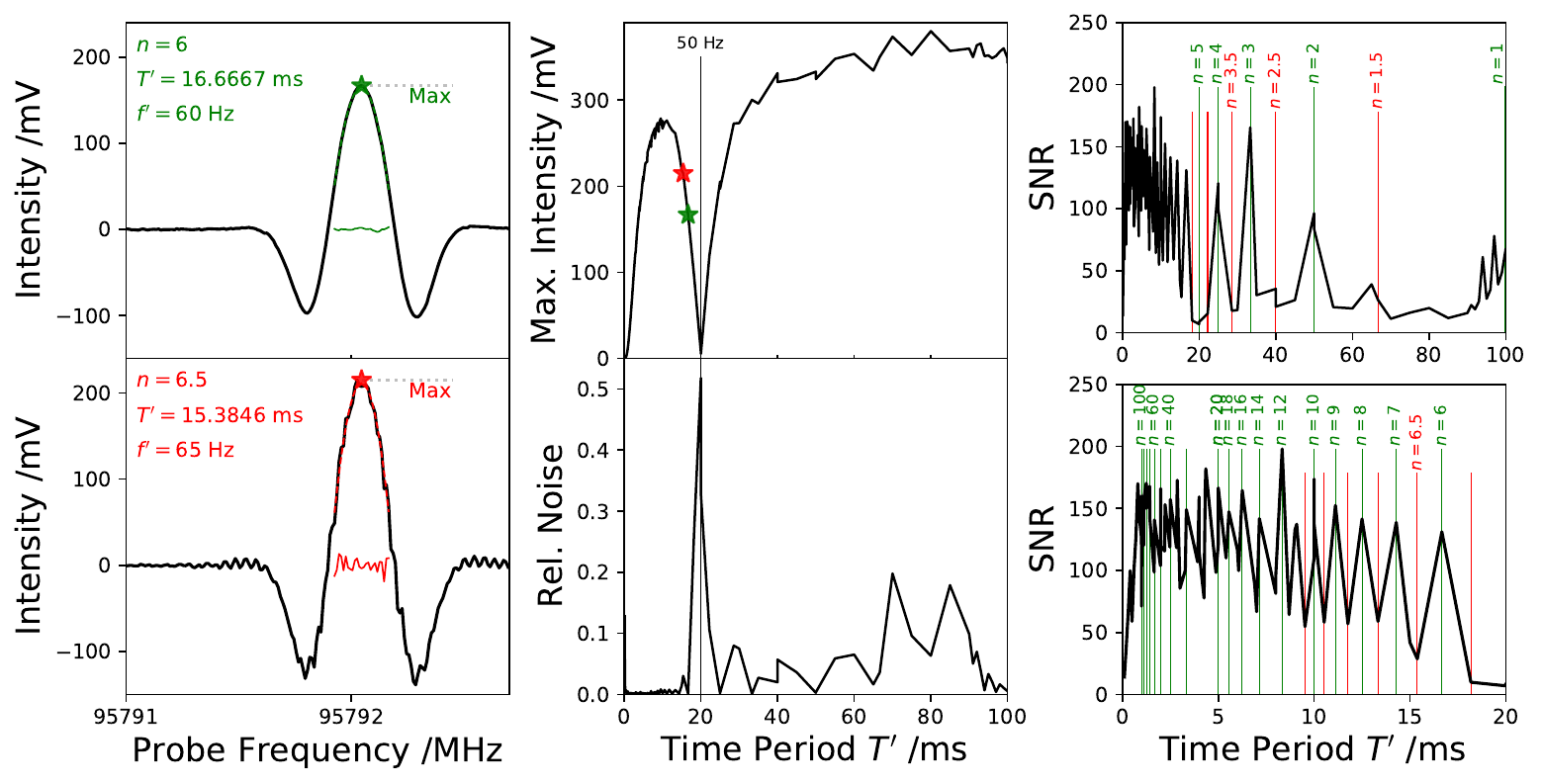}
\caption{
Analysis of the applied modulation frequency of the second lock-in amplifier $f'$ for DM-DR measurements. 
As stated in the main article, the $TC$ should be ideally a multiple $n$ of $T'=1/f'$.
The number of multiples (cycles with exactly one DR on plus one DR off measurement) for $TC=50$\,ms is $n= (2\times TC) / T'= 100\,ms/ T'$, the factor of two is occurring because a 6\,dB/octave filter is used within the lock-in amplifier, which counter-intuitively averages the signal for $2\times TC$.
Two example measurements, for good (green; $n=6$) and bad (red; $n=6.5$) choices are shown on the left hand side.
Half-integers lead to alternation of one more cycle of DR on or DR off for adjacent frequency points, which can be seen in the "zigzag noise" pattern ($n=6.5$; lower left). 
In the center column, the determined maxima and respective noise levels for different time periods with $0.1\leq T'\leq 100$\,ms are plotted.
Here, the effect of a 50\,Hz filter, to get rid of noise from the supply frequency, is clearly seen (center panels).
The determined SNRs are plotted on the right hand side for time periods from 0 to 100\,ms (top) and a zoom from 0 to 20\,ms (bottom).
The determined noise level depends on the chosen frequency window and fit function but the overall interpretation is irrevocably. Integer values of $n$ show better SNRs, but for lower $T'$ this effect becomes smaller. In this work the modulation frequency was chosen to be $T'\approx4.3478$\,ms ($n=23$) with $TC=50$\,ms.
}
\label{FigA:DMDR_Modulationsfrequenz}
\end{figure*}

\begin{figure}[t]
  \centering
\begin{minipage}[c]{.65\linewidth}
  \includegraphics[width=0.9\linewidth]{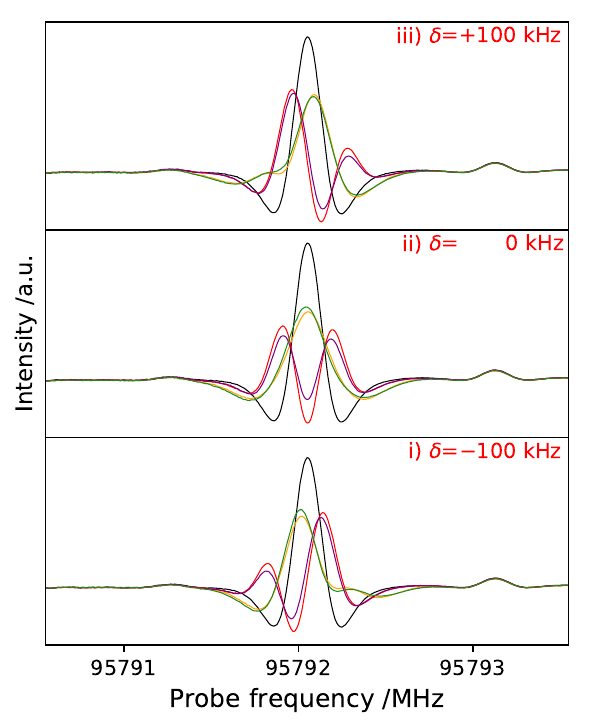}
\end{minipage}
\hspace{0.2cm}
\begin{minipage}[c]{.30\linewidth}
  \includegraphics[width=0.9\linewidth]{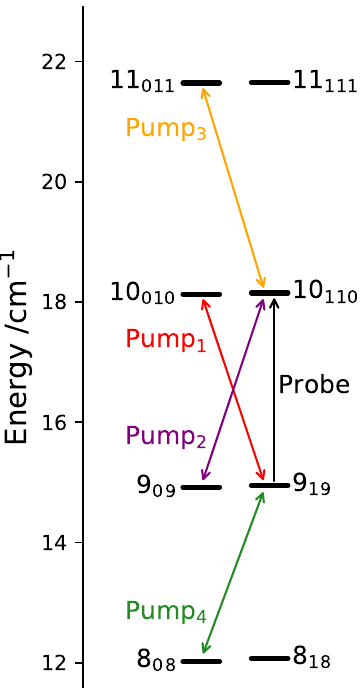}
\end{minipage}
\caption{
DR measurements of \textit{syn}-propanal for all four possible energy level arrangements. 
Regressive energy level arrangements are shown 1) in red ($V$-type) and 2) in violet ($\Lambda$-type) and progressive ones 3) in orange and 4) in green.
A conventional measurement is also shown for comparison (in black).
On the left hand side DR spectra are shown for three different detunings $\delta=\omega-\omega_{21}=+100$\,kHz, $0$\,kHz, $-100$\,kHz.
The probed frequency window is always the same and measured spectra are plotted in the respective color of the pump transition.
Regressive (red and violet) and progressive (orange and green) energy level arrangements can be distinguished if off-resonant pumping is applied.
The more intense Autler-Townes component is red- or blue-shifted in frequency for regressive or progressive arrangements if $\omega>\omega_{21}$ ($\delta=+100$\,kHz), respectively, and vice versa if $\omega<\omega_{21}$ ($\delta=-100$\,kHz). 
Furthermore, a larger splitting can be observed for regressive pumping schemes (cf. Fig.~\ref{Fig:ProReg}). 
The two Autler-Townes components are split but still blended for progressive pumping schemes with on-resonant pumping ($\delta=0$\,kHz). This can be seen by the broader and less intense line shape in comparison to the conventional measurement. 
The origin of different observed splittings is suspected to originate from the substructure (internal rotation components which are blended) and is planned to be described in full detail in the future.
}
\label{FigA:DM_arrangement_measurement}
\end{figure}

Finally, four additional measurement results are presented.
In the first case, the so-called AC Stark shift is determined as a function of the detuned pump frequencies, see Fig.~\ref{FigA:DMDR_PumpOffset}. It is highlighted that a 120\,MHz detuned pump frequency has a similar effect as if the pump power would be turned off.
Furthermore, in Fig.~\ref{FigA:DMDR_Modulationsfrequenz}, the effect of different modulation frequencies of the second lock-in amplifier are analyzed for the SNRs of DM-DR measurements.
The SNRs of DM-DR measurements with regressive arrangements are observed to be larger than progressive ones, therefore DR measurements for various arrangements are compared to a conventional measurement in Fig.~\ref{FigA:DM_arrangement_measurement}.
Last, a broad 2D DM-DR spectrum of ethyl cyanide is shown to demonstrate that even linkages close in frequency can be discriminated.

\bibliographystyle{elsarticle-num-names}
\bibliography{bib}

\end{document}